\documentclass[sigconf]{acmart}
\renewcommand\footnotetextcopyrightpermission[1]{} % Removes copyright footnote

\usepackage{subcaption}
\usepackage{lscape}
\usepackage{xcolor}

\usepackage{balance}
\usepackage{newtxtext,newtxmath}
\usepackage{colortbl}
\usepackage{makecell}    % For multi-line cells
\usepackage{booktabs}    % For better table lines
\usepackage{tabularx}
\usepackage{graphicx}
\usepackage{bmpsize}
\usepackage{textcomp}

\usepackage{comment}
\usepackage{enumitem}
\usepackage{tabularray}
\setlist[itemize]{leftmargin=*}

\setlist[enumerate]{leftmargin=1.15cm}
\usepackage {dsfont}
\usepackage{tablefootnote}
\usepackage{multirow}

\definecolor{darkgreen}{RGB}{0,128,0} % Define dark green color

\newcommand{\fakepar}[1]{\vspace{1mm}\noindent\textbf{#1:}}

\definecolor{maroon}{rgb}{0.988, 0.6, 0.51}
\definecolor{yellow}{rgb}{0.969, 0.89, 0.573}
\definecolor{green}{rgb}{0.663, 0.969, 0.573}

\newcommand{\revised}[1]{\textcolor{black}{#1}}

\begin{document}
%\fancyhead{}
\def\thetitle{Glitch in Time:\\ Exploiting Temporal Misalignment of IMU for Eavesdropping}
% unheard frequencies beyond Nyquist
\title[Glitch in Time]{\thetitle}

\author{Ahmed Najeeb}
%\authornote{Most of the work was completed while the author was a student at LUMS.}
\affiliation{
\institution{LUMS, Pakistan \&\\ Rochester Institute of Technology} 
\country{USA}
}

\author{Abdul Rafay}
\affiliation{
\institution{LUMS}
\country{Pakistan} 
}

\author{Muhammad Hamad Alizai}
\affiliation{
\institution{LUMS}
\country{Pakistan} 
}

\author{Naveed Anwar Bhatti}
\orcid{0000-0003-4115-9889}
\affiliation{
\institution{LUMS}
\country{Pakistan} 
}

\renewcommand{\shortauthors}{Najeeb et al.}

\begin{abstract}
  % Abstractly this is an ACM CCS Template. Keep it short and simple, highlight
  % the main problem and give your punch line contributions. For example,

  % Setting up the ACM CCS template is non-trivial. This is a document to help you
  % get started with ACM CCS template over Overleaf quickly. I also provide some
  % macros in the \texttt{defs.tex} file, that can be helpful for new writers.

% With the rising popularity of voice assistants and other voice-based applications, acoustic eavesdropping attacks have become an increasing concern for users' privacy.  To address this issue, Google imposed a rate limit of 200 Hz on permission-free access to IMUs, rendering these side channel attacks obsolete. Our paper introduces a new exploit that induces temporal misalignment between the gyroscope and accelerometer. We leverage this exploit to propose a novel method called STAG that uses sensor fusion to upsample 200 Hz data and re-enable previously obsolete attacks. When compared with the state-of-the-art, STAG demonstrates a substantial 86\% reduction in word error rate. This underscores the persistent security threat posed by IMUs, even within the confines of a 200 Hz permission-free zone.

The increasing use of voice assistants and related applications has raised significant concerns about the security of Inertial Measurement Units (IMUs) in smartphones. These devices are vulnerable to acoustic eavesdropping attacks, jeopardizing user privacy. In response, Google imposed a rate limit of 200 Hz on permission-free access to IMUs, aiming to neutralize such side-channel attacks. Our research introduces a novel exploit, \textit{STAG}, which circumvents these protections. It induces a temporal misalignment between the gyroscope and accelerometer, cleverly combining their data to resample at higher rates and reviving the potential for eavesdropping attacks previously curtailed by Google's security enhancements. Compared to prior methods, \textit{STAG} achieves an 83.4\% reduction in word error rate, highlighting its effectiveness in exploiting IMU data under restricted access and emphasizing the persistent security risks associated with these sensors.
\end{abstract}

% In such attacks, adversaries exploit a side channel based on zero-permission access to Inertial Measurement Unit (IMU) data to reconstruct speech signals emitted by loudspeakers, potentially compromising users' private information.




\maketitle
% \keywords{LaTeX template, ACM CCS, ACM}

% Section I
\section{Introduction}
\label{sec:intro}

Smartphones' Inertial Measurement Units (IMUs), comprising accelerometers and gyroscopes, are vulnerable to eavesdropping attacks that exploit their capacity to detect vibrations from human speech~\cite{8418650,10.1145/2162081.2162095,184479,10.1145/3448300.3468499,10.1145/2742647.2742658}. While earlier studies focused on reconstructing human voice signals, which would require capturing the entire audio spectrum up to 40 kHz, recent advancements emphasize recognizing patterns within the captured data. 

\emph{Reconstruction}, requiring high-frequency data, aims to recreate the original audio signal, whereas \emph{recognition} focuses on identifying specific speech patterns or features achievable with lower-frequency data. The voiced speech of an adult male typically has a fundamental frequency of 85 to 180 Hz, while that of an adult female ranges from 165 to 255 Hz~\cite{ryalls1982fundamental}. These frequencies fall within the lower range of human speech and can be partially captured by smartphone motion sensors. Despite their limited sampling rate, which typically goes up to 600 Hz, these sensors are capable of detecting key low-frequency speech features, making them a potential target for eavesdropping attacks. Understanding these frequency ranges is crucial, as it underscores the risk that even low-frequency components of speech can be exploited, especially given that fundamental speech frequencies are within the detection capabilities of IMUs.

Earlier efforts like Accessory~\cite{10.1145/2162081.2162095} and Gyrophone~\cite{184479} assessed the feasibility of these attacks, albeit with limited accuracy. Subsequent developments, such as Spearphone~\cite{10.1145/3448300.3468499}, Accelword~\cite{10.1145/2742647.2742658}, Speechless~\cite{8418650}, and StealthyIMU~\cite{inproceedings}, capitalized on the proximity of IMUs to smartphones' loudspeakers, improving the effectiveness of attacks. Nonetheless, security enhancements in Android 12~\cite{androidBehaviorChanges}, which restrict IMU sampling to 200 Hz and require user permission for higher rates, have effectively mitigated many of these zero-permission attack strategies. The reduction of the permission-free zone to 200 Hz has effectively eliminated the majority of low-frequency components, thereby leaving an insufficient amount of information for developing effective recognition systems, which is exceedingly challenging under these constraints. This reflects the dynamic interplay of evolving threats and countermeasures in smartphone security.

\begin{figure}[t]
     \begin{subfigure}[t]{1\columnwidth}
         \centering
         \includegraphics[width=1\linewidth]{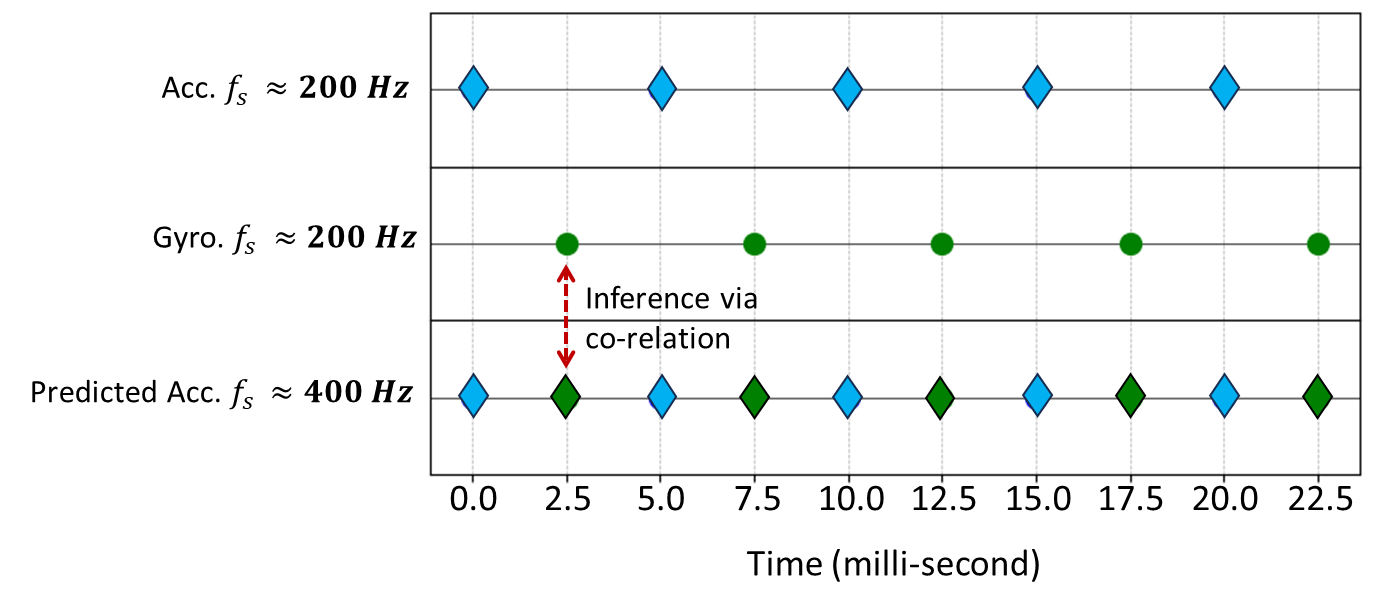}
         \caption{Illustration of optimal temporal misalignment in IMU sensors.}
         \label{fig: misalignment}
     \end{subfigure}
     \begin{subfigure}[t]{1\columnwidth}
         \centering
         \includegraphics[width=1\linewidth]{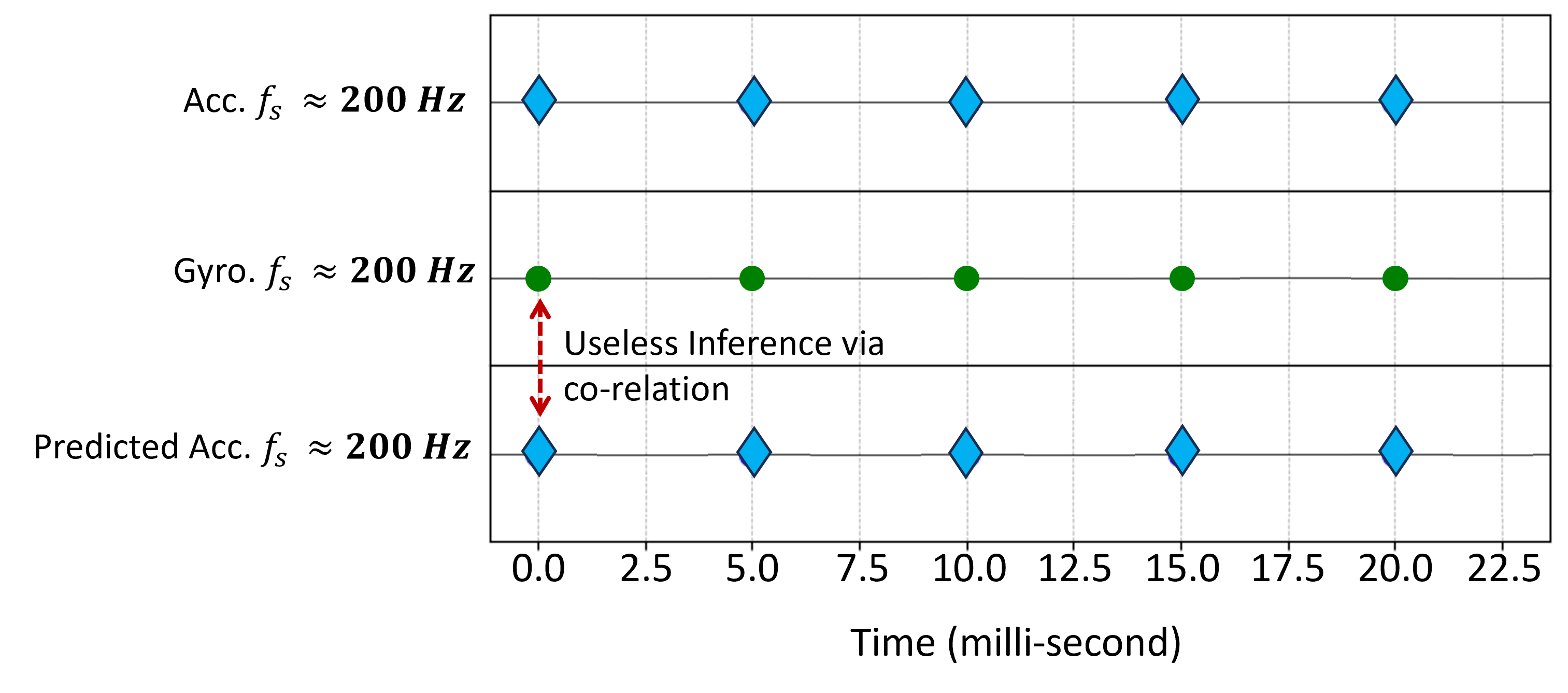}
         \caption{Standard alignment in conventional IMU sensor configurations.}
         \label{fig: alignment}
     \end{subfigure}
     \caption{Data sampling of IMU sensors: contrasting optimal temporal misalignment vs. standard alignment in smartphones.}
        \label{fig:misal intro}
\end{figure}

We introduce \textbf{S}ensor Fusion via \textbf{T}emporal Misalignment in \textbf{A}ccelerometers and \textbf{G}yroscopes (\textbf{\textit{STAG}}), a novel approach that effectively circumvents the 200 Hz sampling rate limitation in modern smartphones. A critical aspect of our research is the introduction of controlled temporal misalignment between accelerometer and gyroscope readings, a technique that reestablishes the eavesdropping threat posed by IMUs. This misalignment leverages the established correlation~\cite{yoon2014improvement,webber2021human} between these data types to enhance accelerometer data upscaling. By deliberately avoiding the synchronization of sensor readings, we prevent data overlap and significantly enhance the accuracy of data fusion. Under conditions where a 200 Hz sampling rate is utilized without special permissions, a temporal misalignment of approximately 2.5 ms between the gyroscope and accelerometer enables the reconstruction of an upsampled 400 Hz signal. This 2.5 ms misalignment is strategically set at the midpoint between two consecutive accelerometer samples, as shown in Figure~\ref{fig: misalignment}. The intentional deviation from standard sensor operation, where accelerometer and gyroscope data are typically aligned in time, distinguishes our method from conventional sensor fusion techniques, as illustrated in Figure~\ref{fig: alignment}.

We have devised a sophisticated data processing pipeline to upscale accelerometer data, incorporating gradient boosting~\cite{Ayyadevara2018} and interpolation~\cite{1456237}. This integration markedly enhances the precision of upsampled data by intelligently combining it with temporally misaligned gyroscope readings. Our method exceeds the performance of existing 200 Hz techniques like InertiEAR~\cite{9796890} and StealthyIMU~\cite{inproceedings}. Unlike InertiEAR, which only recognizes predefined words, our approach supports continuous speech recognition—a significant functional enhancement. Moreover, our method records a dramatically lower Word Error Rate (WER) of \textbf{13\%} compared to StealthIMU's \textbf{78.75\%}, demonstrating superior accuracy in complex speech recognition at reduced sampling rates. This accomplishment represents a significant step forward in understanding zero-permission voice eavesdropping and draws attention to an important security concern in smartphone operating systems. Addressing this vulnerability will strengthen the overall security and safety of Android users, much like how previous efforts to resolve security issues have benefited the broader community.

We make the following contributions:

\begin{enumerate}[label=\textbf{[C\arabic{enumi}]}]
\item {\bf{Demonstrating inadequacy of existing security measures:}} Our research demonstrates that the 200 Hz sampling rate restriction introduced in Android 12 and later versions is inadequate for preventing sophisticated eavesdropping attacks via motion sensors, necessitating a comprehensive reevaluation of security criteria for sensor data in smartphones.

\item {\bf{Temporal misalignment technique:}} We have developed a novel method to induce controlled temporal misalignment in IMU data. This technique strategically adjusts the synchronization between accelerometer and gyroscope readings, which enhances the precision of data fusion processes. It significantly improves the upscaling of accelerometer data from 200 Hz to 400 Hz.

\item {\bf{Upscaling pipeline:}} We have introduced an advanced data processing pipeline that innovatively integrates the Light Gradient Boosting Machine (LightGBM) with interpolation. This approach is tailored to improve audio signal recognition through IMU data, significantly enhancing the accuracy and efficiency of audio recognition at reduced sampling rates.
\end{enumerate}

\section{Preliminaries}
\label{sec: PRELIMINARIES}

We present a fundamental overview of key concepts and the current state of research pertinent to our study. This overview encompasses an examination of MEMS sensors, specifically accelerometers and gyroscopes, a comprehensive review of Gradient Boosting Machines and Sensor Fusion techniques, and an analysis of recent advancements in Spoken Language Understanding. Through this exploration, we lay the theoretical foundation for our research and contextualize our contributions within the landscape of ongoing developments in these areas.

\subsection{Background}
\label{sec:background}
We explore the technical details that play distinct and interrelated roles in advancing our research objectives.

\subsubsection{MEMS Sensors:}
Smartphones are typically equipped with both a three-axis accelerometer and a three-axis gyroscope, which are MEMS~\cite{6495768,maenaka2008mems}. These sensors are essential for detecting device motion, orientation, vibration, and shock. The accelerometer measures acceleration on three axes using a seismic mass, fixed electrodes, and springs. Movements shift the seismic mass, changing the capacitance between electrodes, and converting into acceleration measurements. This setup also detects mechanical vibrations, interpreting them as acceleration. Similarly, the gyroscope measures angular rates using the Coriolis force~\cite{thielman2002proposed}, with a vibrating mass that moves along two axes. External movements induce a Coriolis force that displaces the mass and alters the capacitance, from which angular rates are calculated. In Android smartphones, motion sensor data can be accessed at sampling rates up to 200 Hz without special permissions and up to 600 Hz with permissions, depending on the device's capabilities.

\subsubsection{Gradient Boosting Machines (GBM):}
These are advanced ensemble learning methods employed in machine learning for regression and classification tasks \cite{10.3389/fnbot.2013.00021}. These machines iteratively build an ensemble of weak predictive models, primarily decision trees, to create a robust predictive model. Each subsequent model aims to correct the errors of the previous ones, refining predictions incrementally. The `gradient' in gradient boosting refers to the application of the gradient descent algorithm, which minimizes the model's error. GBMs are noted for their predictive accuracy, adaptability to various data types, and robustness against overfitting. Our research utilizes LightGBM~\cite{NIPS2017_6449f44a}, a GBM variant known for its computational efficiency and minimal memory usage, employing a histogram-based decision tree learning algorithm suitable for limited-resource environments.

\subsubsection{Sensor Fusion:}
It is an advanced data processing technique that combines data from multiple sensors to achieve a more accurate and comprehensive understanding than is possible with individual sensors alone \cite{SASIADEK2002203,sasiadek2002sensor,elmenreich2002introduction}. This technique is critical in applications where integrating diverse data sources is vital to enhanced perception and decision-making. It merges data from accelerometers, gyroscopes, video cameras, and WiFi signals, improving system accuracy and reliability. Sensor fusion is instrumental in navigation systems like Google Maps, where it integrates GPS and sensor data to precisely determine a user's location and orientation \cite{googleManageYour}. In our research, we apply sensor fusion principles to effectively combine accelerometer and gyroscope data using gradient boosting, aiming to refine data sets for improved eavesdropping attack models.

\subsubsection{Spoken Language Understanding (SLU):}
Voice assistants must interpret and respond to human speech~\cite{1511821}. SLU processes involve intent detection, classifying spoken phrases according to their purposes, and slot filling, which extracts details like dates and locations from speech. Modern SLU systems may use a cascaded pipeline that first transcribes speech to text before applying Natural Language Understanding (NLU), or an end-to-end model that directly maps speech signals to intents and slots, thereby avoiding transcription errors \cite{inproceedings}. Recent advancements have introduced joint models that manage both tasks simultaneously and incorporate pre-trained language models and deep neural networks, significantly enhancing SLU performance. Our study utilizes StealthyIMU \cite{inproceedings} SLU as a benchmark, an end-to-end SLU model proficient at extracting sensitive data from motion sensor signals, illustrating this method's efficacy in utilizing sensor-based data leakage.

% \begin{table}[t]
% \centering
% \caption{Overview of related work.}
% {
% \begin{tabular}{p{3cm}lp{4cm}ll}
% \toprule
% \textbf{Paper}  & \textbf{Sensor} & \textbf{Recognition} & \textbf{Evaluation Metric} & \textbf{Accuracy \newline @ 200 Hz} \\
% \midrule
% Gyrophone \cite{184479} & Gyro & Digits (\textcolor{red}{\bfseries constrained}) & Accuracy   &  17\% \\
% \midrule
% Accelword \cite{10.1145/2742647.2742658} & Acc    & Words (\textcolor{red}{\bfseries constrained})    & F-measure &  85\%                      \\
% \midrule
% PitchIn \cite{7944789}  & Acc    & Words \newline(\textcolor{red}{\bfseries constrained})       & Accuracy =                      & NA                      \\
% \midrule
% Spearphone \cite{10.1145/3448300.3468499} &
%   Acc &
%   Digits + Words \newline(\textcolor{red}{\bfseries constrained}) &
%   F-measure &
%   NA \\
% \midrule
% AccelEve \cite{inproceedings-acceleve}& Acc & Digits + Letters (\textcolor{red}{\bfseries constrained}) & Accuracy &  NA \\
% \midrule
% AccEar \cite{9833716}  & Acc    & Words (unconstrained)      & 1-WER                          & NA                      \\
% \midrule
% InertiEar \cite{9796890} & Acc+Gyro & Digits + Letters (\textcolor{red}{\bfseries constrained}) & Accuracy  &78.8\%\\
% \midrule
% StealthyIMU \cite{inproceedings} & Acc & Spoken Language Understanding \newline(unconstrained) & 1-SER \% &37\%\\
% \midrule
% \rowcolor{green} STAG [Ours] & Acc+Gyro & Spoken Language Understanding \newline(unconstrained) & 1-SER & 58\% \\

% \bottomrule
% \end{tabular}
% }
% \label{tab:relw}
% \end{table}

\begin{table*}[t]
\centering

\caption{Overview of related work}

\begin{tabular}{llm{2.5in}m{1in}m{1in}}
\hline
\rowcolor{lightgray}
\textbf{Paper} &
  \textbf{Sensor} &
  \textbf{Recognition} &
  \textbf{Evaluation Metric} &
  \textbf{Accuracy @ 200 Hz} \\ \hline
Gyrophone \cite{184479} &
  Gyro &
  Digits (\textcolor{red}{\bfseries constrained}) &
  Accuracy &
  17\% \\ \hline
Accelword \cite{10.1145/2742647.2742658} &
  Acc &
  Words (\textcolor{red}{\bfseries constrained}) &
  F-measure &
  85\% \\ \hline
PitchIn \cite{7944789} &
  Acc &
  Words (\textcolor{red}{\bfseries constrained}) &
  Accuracy &
  NA \\ \hline
Spearphone \cite{10.1145/3448300.3468499} &
  Acc &
  Digits + Words (\textcolor{red}{\bfseries constrained}) &
  F-measure &
  NA \\ \hline
AccelEve \cite{inproceedings-acceleve} &
  Acc &
  Digits + Letters (\textcolor{red}{\bfseries constrained}) &
  Accuracy &
  NA \\ \hline
AccEar \cite{9833716} &
  Acc &
  Words (unconstrained) &
  1-WER &
  NA \\ \hline
InertiEar \cite{9796890} &
  Acc+Gyro &
  Digits + Letters (\textcolor{red}{\bfseries constrained}) &
  Accuracy &
  78.8\% \\ \hline
StealthyIMU \cite{inproceedings} &
  Acc &
  Spoken Language Understanding (unconstrained) & \begin{tabular}[c]{@{}l@{}}1-SER\\ 1-WER\end{tabular} 
   &
  \begin{tabular}[c]{@{}l@{}}37\%\\ 21.25\% \end{tabular} \\ \hline
\rowcolor{green} STAG [Ours] &
  Acc+Gyro &
  Spoken Language Understanding (unconstrained) &
  \begin{tabular}[c]{@{}l@{}}1-SER\\ 1-WER\end{tabular}  &
  \begin{tabular}[c]{@{}l@{}}58\%\\ 86.98\% \end{tabular}\\ \hline
\end{tabular}

\label{tab:relw}
\end{table*}

\subsection{Related Work}
\label{sec:relwork}
% In~\secref{sec:intro} you talked about the project at a very high-level. This is
% the section from where you will start giving details. First with things that are
% already done, and familiarize the reader with background information they will
% need to understand you work. 

Research on speech recognition using motion sensors broadly falls into two categories. The first involves studies on recognizing external voice inputs through smartphone motion sensors. The second focuses on speech recognition from internal sources, typically using the smartphone's loudspeakers. Table~\ref{tab:relw} presents a detailed comparison of these studies with our approach, STAG. We assess the accuracy of previous methods based on non-permission IMU data sampled below 200 Hz. We further show evaluation metrics used and whether the papers use constrained or unconstrained data. Constrained, in our case, means that the data used is based on a limited set of digits or words, while unconstrained means that the Model can account for unseen vocabulary. 

Gyrophone~\cite{184479} investigates a smartphone placed near a loudspeaker on a solid surface. Using the gyroscope, it detects speech signals from the external loudspeaker for speech recognition and speaker identification. However, the gyroscope's limited sensitivity to surface vibrations and its 200 Hz sampling rate cap pose challenges in achieving high accuracy in recognition tasks.

AccelWorld~\cite{10.1145/2742647.2742658} explores scenarios where a user speaks near a smartphone, either handheld or on a desk. The accelerometer captures speech signals through air transmission, aiding in hot word detection like "Okay, Google" or "Hi, Galaxy." PitchIn~\cite{7944789} uses a distributed time-integrated analog-digital conversion (TIADC) system to approximate high sampling frequencies with low per-node rates. By pooling data from multiple devices, PitchIn demonstrates effective speech recognition, showcasing the potential of eavesdropping through closely-placed devices.

These initial methods primarily utilized external human speech. A more concerning attack vector involves the smartphone's built-in loudspeaker. Since the motion sensors are physically close to the speaker within the same device, they pose a higher risk of revealing sensitive information during calls or voice assistant interactions. Background applications can exploit zero-permission motion sensors to capture and reconstruct this data.

% Spearphone~\cite{10.1145/3448300.3468499} marks the first study exploiting loudspeaker proximity to IMUs. It uses altered accelerometer readings and machine learning to classify gender (over 90\% accuracy) and identify speakers (over 80\% accuracy). Spearphone can reconstruct speech with a word detection accuracy of 67\%. AccelEve \cite{inproceedings-acceleve} employs a deep learning model (DenseNet) to reconstruct speech from accelerometer signal spectrograms, achieving a top-1 accuracy of 78\% for digits and 55\% for digits plus letters.

AccelEve~\cite{inproceedings-acceleve} marks the first study exploiting loudspeaker proximity to IMUs. It employs a deep learning model (DenseNet) to reconstruct speech from accelerometer signal spectrograms, achieving a top-1 accuracy of 78\% for digits and 55\% for digits plus letters. Spearphone~\cite{10.1145/3448300.3468499} utilizes altered accelerometer readings and machine learning techniques to classify gender with over 90\% accuracy and identify speakers with over 80\% accuracy. Additionally, Spearphone can reconstruct speech with a word detection accuracy of 67\%.

AccEar~\cite{9833716} expands the scope to include a wider range of vocabulary. Using a conditional Generative Adversarial Network (cGAN), AccEar enhances spectrograms from low-frequency accelerometer signals to reconstruct high-fidelity audio, achieving 90\% accuracy under optimal conditions. StealthyIMU \cite{inproceedings} extends this to voice assistants, accessing sensitive information like GPS and contact data via an accelerometer. It formulates this side-channel attack as an end-to-end private speech understanding challenge, using Sentence Error Rate (SER) as the metric. SER considers a response to be error-free only if all entities in a single voice assistant response are correctly identified. StealthyIMU reaches an 86\% sentence accuracy (1-SER) and 37\% without permission access, although our evaluation shows a significantly lower sentence detection accuracy.

Google's sampling rate restrictions have limited the effectiveness of these methods, with most underperforming below 200 Hz. InertiEar \cite{9796890} attempts to circumvent this by using IMUs to eavesdrop on both top and bottom smartphone speakers. By exploiting the coherence between accelerometer and gyroscope responses and their hardware diversity, InertiEar achieves a greater than 75\% speech recognition accuracy for digits and letters.

Our study diverges from the existing literature in two crucial aspects. First, we address the sampling rate issue by upsampling the accelerometer signal, thereby enhancing the input quality for existing models. Second, we introduce a novel method to induce perfect temporal misalignment between accelerometer and gyroscope values, maximizing the benefits of data fusion. The closest related work, InertiEar, does not introduce any form of misalignment, which partly explains why our approach achieves higher accuracy (Section \ref{sec:results}). Our contribution, STAG, functions as an effective preprocessing layer for existing ML models, circumventing Google's rate limitations and significantly enhancing the effectiveness of previous eavesdropping attacks.

\section{Problem Formulation}
\label{sec: Problem Formula}
Our research focuses on uncovering a critical vulnerability in the Android operating system, specifically related to how its scheduler impacts the alignment of sensor data streams. Rather than emphasizing application development, we draw attention to a security weakness in synchronizing Android’s sensor data. Misaligned sensor data streams can be exploited for covert surveillance, posing severe privacy risks. Given Android’s status as one of the most widely used operating systems, our findings demonstrate how existing security measures—such as the 200 Hz sampling limit—may create a false sense of protection.

\subsection{Threat Model}
% !!!!!!!!!!!!!!!!!!!!!!!!!!!!!!!!!!!!!!!!!!!!!!!!!!!!!!!!!!!!!!!!!!!!!!!1!!!!
% the extra part is written to cover up the efficiency given by reviewer !!!!
% !!!!!!!!!!!!!!!!!!!!!!!!!!!!!!!!!!!!!!!!!!!!!!!!!!!!!!!!!!!!!!!!!!!!!!!!!!!!
Our research focuses on a typical scenario involving Android smartphone users, who represent a significant segment of the global smartphone market. The relevance of our study is amplified by Android's recent security update in Android 12 in 2021 \cite{androidBehaviorChanges}, which limits the IMU sampling rate to 200 Hz. We examine the potential risks when smartphones broadcast speech signals through their loudspeakers, potentially revealing sensitive personal information. The associated threat escalates due to the extensive permissions often granted to voice assistant applications, such as access to calendar, contacts, location, microphone, phone, SMS, and storage \cite{googleManageYour}, each a potential repository of private data.

Our method, STAG, is designed to harness the potential of zero-permission motion sensors to capture vibrations from various audio sources, including voice assistants and human conversations, even within the constraints of the new sampling rate limits. This is achieved by deliberately inducing temporal misalignment between accelerometer and gyroscope readings. Our threat model envisions a scenario where a user unwittingly installs an app containing STAG. The app could store the harvested data locally or, if permissions allow, upload it to a cloud server for advanced analysis. Popular mobile browsers like Chrome, Brave, and Firefox provide default access to motion sensors without explicit user permission~\cite{sensor-js-2018}. Thus, in addition to embedding itself in standard applications, malicious software can also be deployed through executable scripts (e.g., PyScript, Javascript) on seemingly benign websites, enabling it to process the data locally and surreptitiously upload it to a remote server without the user's knowledge.

\subsection{Attack Scenarios}
% \label{sec: Attack scenarios}
In the scenario of a malicious app installed on a smartphone, the following types of information could be at risk:

\begin{enumerate}[label=\textbf{[AS\arabic{enumi}]},leftmargin=2.6em]
\item {\bf{Location information:}} The malicious app can use responses from the voice assistant, especially for weather-related queries, to infer the user's coarse location, such as their city or county (city district) of residence. The app can pinpoint the general area where the user is located by analyzing the specific details mentioned in the loudspeaker's output.

\item {\textbf{User routine:}} The app leverages SLU to decode responses from the voice assistant to user queries. It systematically collects and analyzes this data to detail the user’s daily routine. Information from reminders and to-do lists reveals scheduled activities, while queries about directions show regular travel routes and frequented locations. This thorough analysis not only maps typical daily activities but also predicts future behaviors based on recurring patterns.

\item {\textbf{Communication sphere:}} Voice assistants, which access users' contact lists for initiating calls and sending messages via voice commands, present a privacy risk when exploited by malicious apps. Such apps can covertly listen to interactions to discern contact identities and capture content confirmed verbally by the voice assistant. This unauthorized access to private message contents could lead to substantial privacy breaches.
\end{enumerate}

These scenarios highlight the diverse ways in which a sophisticated malicious app could exploit the functionalities of a smartphone to extract sensitive personal information, underscoring the need for robust security measures.

\subsection{\revised{End-to-End Attack}}
\begin{figure}
    \includegraphics[width=\linewidth]{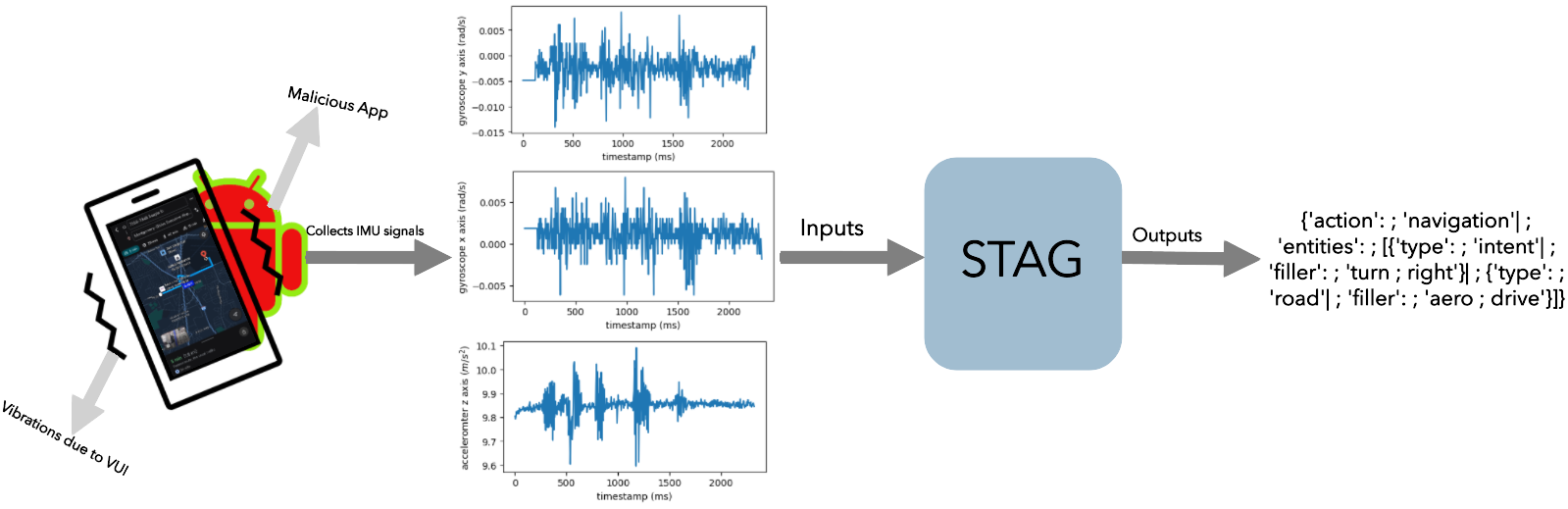}
    \caption{\revised{An end-to-end attack scenario: a malicious app captures IMU data from a navigation app and processes it via STAG to infer sensitive user information.}}
    \label{fig:endtoend}
\end{figure}

\revised{An example of an end-to-end attack is depicted in Fig. \ref{fig:endtoend}. In this scenario, a malicious application runs covertly in the background of a navigation app, continuously capturing IMU data without the user’s knowledge. This data is subsequently processed using the STAG framework to generate a text file containing semantic frames, which are then stored on the device. These semantic frames enable the extraction of intents and entities from each VUI response. By aggregating multiple responses over time, the attacker can infer sensitive user information—such as locations, routes, and home or work addresses—posing a significant threat to user privacy.}
%%%%%%%%%%%%%%%%%%%%%%%%%%%%%%%%%%%%%%%%%%%%%%%%%%%%%%%%%%%%%%%%%%%%%%%%%%%%%%%%

\section{Design and Implementation}
\label{sec:methodology}

STAG has been refined into a robust system capable of augmenting the utility of sensor data for various applications. In the following subsections, we dissect the key components of STAG, detailing the architecture’s complexities and the rationale behind each design decision.

\subsection{Temporal misalignment}
\label{sec:temporal}

To counteract the challenge posed by Android's 200 Hz sampling rate cap on accelerometer sensors, our research undertook a methodical investigation to acquire or infer missing accelerometer data. This investigation was structured around experimenting with three distinct scenarios, all centered on the concept of \emph{temporal misalignment}—a strategic offset in the timing of different sensor readings. The initial scenarios offered valuable, albeit incremental, insights into the problem. However, it was the exploration of the final scenario that led to a breakthrough. Each scenario was a step forward in an iterative process, gradually leading us to the successful implementation of temporal misalignment as a viable solution.

% \begin{figure}[!htbp]
%     \centering
%     \includegraphics[width=\linewidth]{images/after-mag.png}
%     \caption{Effect of introduction of the magnetometer to the samples. The samples are closer to the middle.}
%     \label{fig:after-mag}
% \end{figure}

\subsubsection{Scenario 1: multiple accelerometer instances.}

In our initial approach, we implemented two separate software instances of the accelerometer sensor within our Android application. Each was set up to sample at an approximate rate of 200 Hz, with a slight offset in their initialization times. The intention was to merge these sets of data to simulate an effectively higher sampling rate.

\revised{Although the accelerometer instances had staggered start times, both produced identical timestamps. This occurs because Android’s Sensor Service, which mediates data from the physical sensor via the Sensor HAL, distributes the same data stream, including timestamps from the system’s monotonic clock, to all requesting apps. Consequently, each accelerometer instance receives identical sensor readings. Moreover, Android does not strictly maintain a 200 Hz sampling rate, allowing the Sensor Service to throttle data to manage power and load. As a result, both instances effectively captured the same data.}

\noindent
\emph{\textbf{Lesson Learned}:
Attempting to create multiple software instances of the accelerometer within an Android application to simulate a higher sampling rate is ineffective. \revised{In the Android operating system, the Sensor Service and HAL deliver sensor data in a manner that yields identical timestamps for readings from the same physical sensor, regardless of how many instances request them. This underscores inherent limitations in the OS and hardware design, illustrating that timestamps reflect system-level data handling rather than the precise moment of physical sensor capture.}}

\subsubsection{Scenario 2: gyroscope and accelerometer instances.}

In our subsequent experiment, we implemented a strategy that involved registering both a gyroscope and an accelerometer in the Android application. Each sensor was configured to sample at approximately 200 Hz, with a deliberately introduced delay in their initiation. However, similar to our first attempt, this method also led to both sensors providing readings with identical timestamps.

The cause of this synchronized timestamp phenomenon can be traced back to two main factors. Firstly, Android's management of sensor sampling rates is inherently flexible; while a maximum rate is specifiable, the actual sampling frequency can vary and is adjusted in real-time by the operating system \revised{through the Sensor Service}. This variability in the sampling intervals could result in the gyroscope and accelerometer inadvertently aligning their readings, despite our efforts to stagger their activation. Secondly, the design of IMU sensors, which encompasses both accelerometers and gyroscopes, is inherently inclined towards synchronized functioning. These sensors are typically integrated at both hardware and software levels to facilitate efficient and precise motion detection, which naturally leads to their simultaneous data sampling and reporting. \revised{The Sensor HAL, which interfaces directly with the IMU, often timestamps data from both sensors at the same time, reflecting the synchronous operation of the IMU. The Sensor Service then delivers this data to applications without altering the timestamps. A key finding was the consistent zero variance in sampling delays, indicating that the timestamps were likely being assigned by a software component (either the HAL or the Sensor Service) based on a timer, rather than reflecting the actual hardware capture time.}

\noindent
\emph{\textbf{Lesson Learned}:
Introducing gyroscope and accelerometer sensors with a staggered initialization also fails to achieve desynchronized readings, as the Android OS manages sensor sampling rates flexibly \revised{through the Sensor Service and the Sensor HAL, and IMU sensors are often designed to operate synchronously}. This experiment underlines the challenge of achieving temporal misalignment through software alone, due to the inherent synchronized functioning of IMU components, \revised{the HAL's timestamping practices}, and the OS's real-time adjustments to sampling rates}

\subsubsection{Scenario 3: introducing magnetometer.}
A pivotal discovery emerged during the integration of magnetometer with gyroscope and accelerometer systems: the unexpected desynchronization of their readings, which enhanced the effectiveness of sensor fusion. We achieved the desired misalignment by enforcing consecutive gyroscope and accelerometer readings to be interleaved with magnetometer readings.

The distinct hardware configurations of sensors, particularly the STMicroelectronics LSM6DSM IMU~\cite{lsm6d} found in the Samsung A31/A33 and the Bosch BMI160 IMU~\cite{boschsensortec} used in the Pixel 5, involve direct connections between the IMU and the host, with the magnetometer interfacing through the IMU. \revised{Hence, the magnetometer lacks a dedicated communication channel to the host processor and instead communicates through the IMU, often via an I2C interface.}. This architectural design involves a FIFO buffer within the IMU that aids in efficient high-frequency data capture and energy savings. However, it can introduce synchronization problems when integrating a Magnetometer that does not have its own FIFO buffer and whose values are stored in the FIFO of IMU—leading to data misalignments, as our research indicates. \revised{Specifically, when the magnetometer is enabled, its data is queued into the IMU’s FIFO buffer along with the accelerometer and gyroscope readings. This extra data flow disrupts the expected timing of accelerometer and gyroscope sample retrieval in the Sensor HAL. For example, the HAL might read an accelerometer sample, then unexpectedly encounter magnetometer data, followed by a gyroscope sample, altering the timing relationship between the accelerometer and gyroscope data as observed by the Sensor Service and application. The exact disruption mechanism varies by IMU and HAL implementation, often involving the HAL reading more data than anticipated or experiencing shifted interrupt timings due to the added magnetometer entries.}

\noindent
\emph{\textbf{Lesson Learned}:
Integrating a magnetometer with gyroscope and accelerometer systems can induce the desired temporal misalignment, enhancing sensor fusion. The magnetometer’s subordinate role to the IMU and the presence of FIFO buffers drive this desynchronization. Adding magnetometer data to the IMU’s FIFO disrupts the expected sequence and timing of accelerometer and gyroscope data retrieval by the HAL, resulting in observable misalignment. This interaction can be leveraged to improve recognition systems, but it also exposes security vulnerabilities in some smartphone models, where the misalignment may be exploited for unauthorized data extraction.}

\begin{figure}
    \centering
         \includegraphics[width=\columnwidth]{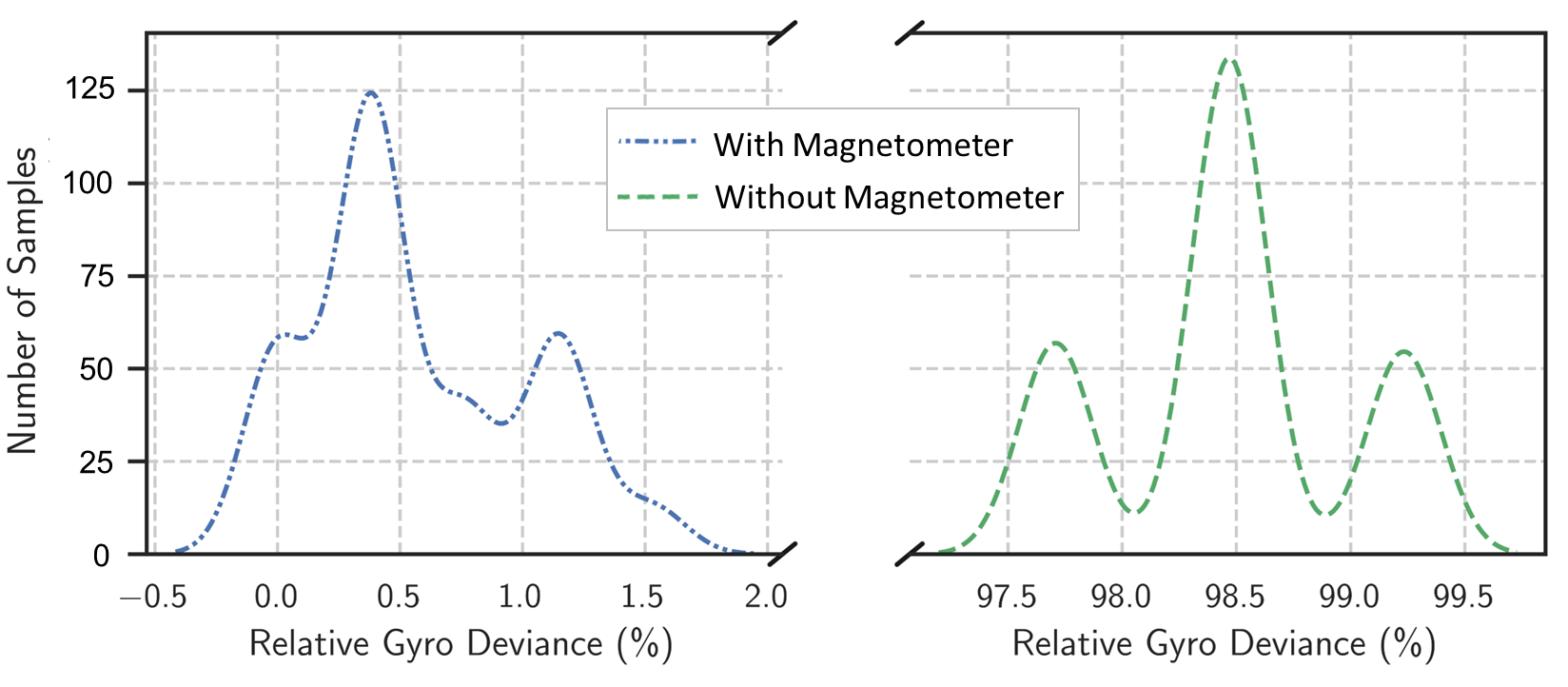}
        \caption{Temporal misalignment and sample distribution in the Samsung Galaxy A33 before and after integrating the magnetometer.}
        \label{fig:misal}
\end{figure}

Table~\ref{tab:devices} classifies devices based on gyroscope deviation from the center, with or without a magnetometer. Devices in red show unchanged deviation; yellow devices, like the ONEPLUS 6T, display minimal deviation; green devices significantly approach the ideal misalignment. 
Despite optimal alignment at 2.5 ms, data fusion accuracy remains consistent with up to a 25\% deviation, demonstrating the complex nature of current smartphone sensor systems. Figure~\ref{fig:misal} for the Samsung Galaxy A33 illustrates how gyro deviation consolidates into the `ideal region' post-magnetometer integration, emphasizing substantial shifts toward desired misalignment.

\begin{table}
\centering
\caption{Gyroscope deviation from the ideal central position, with and without magnetometer influence, across various smartphone models.}
\begin{tabular}{lcc}
\toprule
{\multirow{2}{*}{Device}} &
  
  \multicolumn{2}{c}{Gyro deviation from centre (\%)} \\ \cmidrule(l){2-3} 
  &
  {No Mag} &
  {With Mag} \\ \midrule
\rowcolor{maroon}Redmi Note 12         & 100.0000   & 100.0000   \\
\rowcolor{maroon}Samsung Galaxy A73        & 100.0000   & 100.0000   \\
\rowcolor{yellow}ONEPLUS 6T                 & 0.0093     & 0.0092     \\
\rowcolor{green} Samsung Galaxy A33         & 98.4629    & 0.5662     \\
\rowcolor{green} Samsung Galaxy M31         & 98.7281    &  0.5797  \\ 
\rowcolor{green} Google Pixel 5          & 51.4633    & 22.9684    \\
\bottomrule
\end{tabular}
\label{tab:devices}
\end{table}

These results show that temporal misalignment can be exploited to increase the effective sampling rate. After achieving misalignment between the accelerometer and gyroscope readings, each operating at 200 Hz with a 2.5 ms offset, we successfully doubled the effective sampling rate to 400 Hz. However, further upsampling beyond this limit would only rely on repeating or interpolating existing data, leading to diminishing returns in quality and accuracy. Therefore, our analysis focused on this limit to demonstrate the efficacy of our method within the Android system's constraints.

% Is it OK now? I added extra part in brackets.
% IMUs have a built-in FIFO buffer that enables them to store samples, allowing for a higher sample rate and decreasing the host MCU interaction with the sensor, which allows system power savings.  Magenetometers don't tyup

% This configuration likely contributes to the synchronization issues observed in our experiments.

\begin{figure}
\centering
\includegraphics[width=0.9\columnwidth]{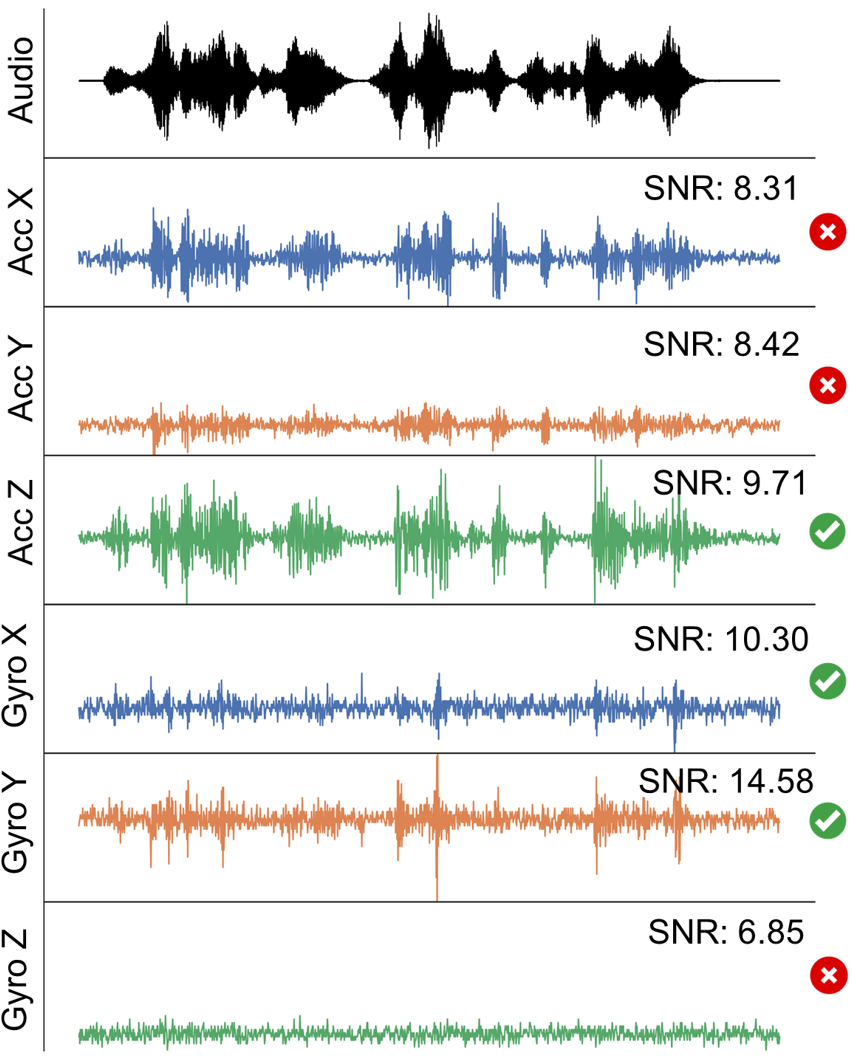}
\caption{Normalized accelerometer and gyroscope responses to audio. The y-axis is scaled from 0 to 1. The accelerometer shows a stronger response than the gyroscope, but both sensors exhibit discernible patterns.}
\label{fig:Acc-gyro-waves}
\end{figure}

\subsection{Establishing Correlation}
\label{subsec: Corr}
To fully leverage the capabilities of STAG's temporal misalignment, a critical analysis of both the signal-to-noise ratio (SNR) and the correlation index between accelerometer and gyroscope data is imperative. The SNR provides an indication of the quality of the signal, whereas the correlation index reveals the strength of the relationship between the two data streams, a connection that is not inherently ensured by a high SNR alone.

For our purposes, the Z-axis of the accelerometer is consistently selected due to its superior SNR, which remains stable across different smartphone positions and varying sound volumes. This stability is primarily attributed to its alignment with the loudspeaker vibrations, as depicted in Figure \ref{fig:Acc-gyro-waves}. However, a high SNR in the accelerometer does not automatically imply a strong correlation with the gyroscope's data. Therefore, assessing the correlation index is equally essential to ascertain that the chosen axes of the gyroscope exhibit a significant relationship with the accelerometer’s Z-axis, a crucial factor for precise data prediction.

\begin{figure}
\centering
\includegraphics[width=\columnwidth]{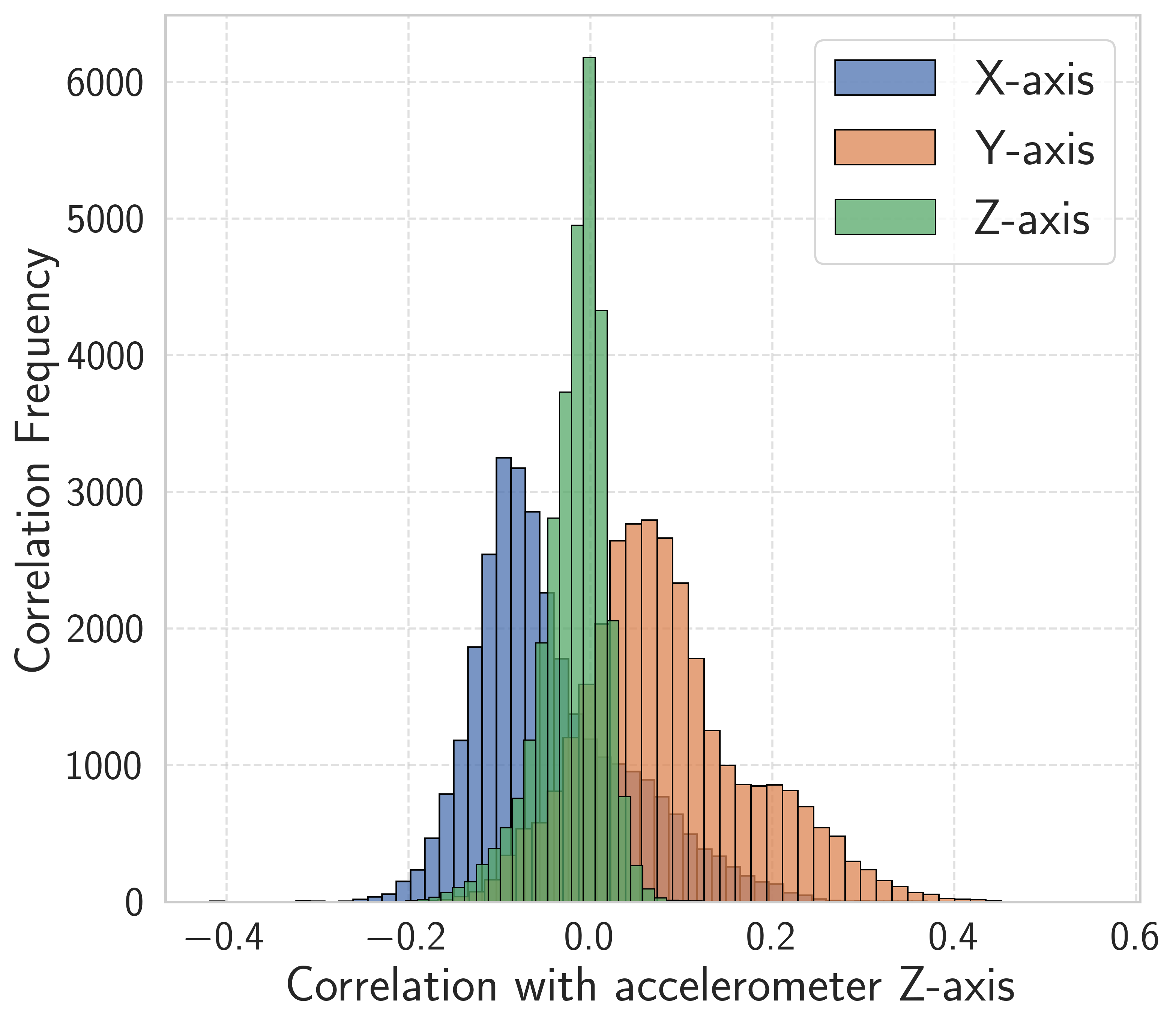}
\caption{Gyroscope axes correlation with Z-axis of the accelerometer.}
\label{fig:gyro-corr}
\end{figure}

Our investigation indicated that while the accelerometer's Z-axis showcased the highest SNR, it was the X and Y axes of the gyroscope that exhibited the most substantial correlation with the Z-axis of the accelerometer, as evidenced in Figure \ref{fig:gyro-corr}. While this correlation figure is specific to the Samsung Galaxy A33, similar patterns were observed in all the mobiles mentioned in Table~\ref{tab:devices}, reinforcing the generalizability of our findings. This finding implies that, regardless of their SNR, the gyroscope's measurements on the X and Y axes are effectively correlated with the linear acceleration detected by the accelerometer’s Z-axis. Consequently, these axes become ideal candidates for sensor fusion within the STAG framework.

In summary, by choosing the accelerometer’s Z-axis for its high SNR and the gyroscope’s X and Y axes for their strong correlation with the accelerometer’s Z-axis, we equip STAG’s machine learning models with data that is of high quality and strongly correlated. This strategic combination of SNR and correlation index is essential in successfully predicting missing accelerometer data and achieving precise sensor fusion in STAG’s architecture.

\subsubsection{\revised{Physical Basis for Sensor Correlation}}
\revised{The observed correlation between the accelerometer’s Z-axis and the gyroscope’s X and Y-axes arises from basic motion principles. When a smartphone accelerates linearly along its Z-axis (e.g., moving vertically), inertial forces induce slight rotations around the X and Y axes. Essentially, changes in linear acceleration lead to corresponding changes in angular velocity. Moreover, precise IMU design and calibration ensure that these measured correlations reflect actual motion dynamics rather than sensor crosstalk or electromagnetic noise. Consequently, gyroscope data can reliably indicate positional changes and serve as a basis for predicting future positions.}

% \begin{figure}[!htpb]
%     \centering
%     \includegraphics[width=0.4\linewidth]{images/correlation_plot.png}
%     \caption{Gyroscope axes correlation with Z-axis of the accelerometer.}
%     \label{fig:gyro-corr}
% \end{figure}

\subsection{Data collection}
\label{sed:Data collection}
    
% To enhance the performance of our STAG system in interpreting speech labels, we collected over 16,000 new sensor readings from accelerometers and gyroscopes. This was in addition to utilizing 20,000 existing readings from the StealthyIMU dataset~\cite{stealthyimu_vui_dataset}. For textual data, we employed Google's Directions API to obtain travel routes between a combination of over 100 cities and famous landmarks. These routes were then converted into spoken words using Google's Text-to-Speech service, which leverages the WaveNet model—the same technology used by Google's voice assistant \cite{googleTypesVoices}. We used SpaCy \cite{spacy}, a python library for natural language processing and regular expression (Regex) techniques to extract all SLU entities mentioned in each trace, ensuring our system accurately interpreted the speech labels.

% To acquire the sensor readings, we developed a simple Android application. This app played spoken directions from .wav files through a smartphone's speaker while simultaneously recording data from the accelerometer and gyroscope at 400 Hz. The readings were subsequently transmitted back to the computer for analysis.

To enhance the performance of our STAG system in interpreting speech labels, we collected over 16,000 new sensor readings from accelerometers and gyroscopes. This was in addition to utilizing 20,000 existing readings from the StealthyIMU dataset~\cite{stealthyimu_vui_dataset}. 

For textual data, we utilized Google's Directions API to derive travel routes between a combination of over 100 cities and prominent landmarks within the United States, ensuring consistency with the StealthyIMU dataset and aiming to enhance prediction accuracy. The city list was generated using GPT and validated through the Maps API. These routes were converted into spoken instructions using Google's Text-to-Speech service, which employs the WaveNet model—technology also utilized by Google's voice assistant~\cite{googleTypesVoices}. To extract all SLU entities from each route description, we employed SpaCy~\cite{spacy}, a Python library for natural language processing, and combined it with regular expression techniques and the GPT-4 API. This rigorous process ensures that our system accurately interprets the speech labels. All collected data, including each wave file and associated metadata, were organized and stored in CSV format.

\begin{figure*}
    \includegraphics[width=\textwidth]{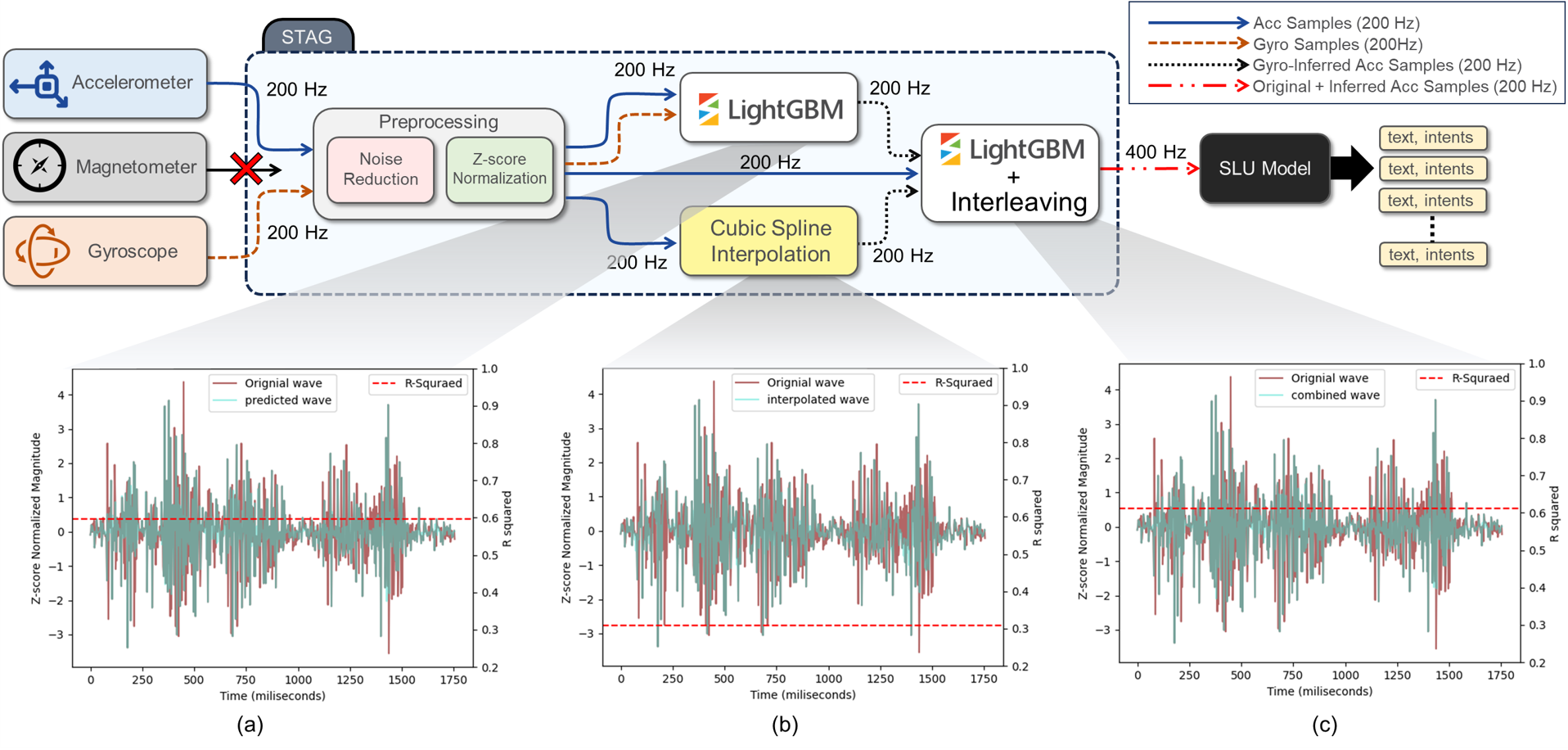}
    \caption{Overview of the STAG architecture and its accuracy assessment. \textbf{(a)} LightGBM predictions on normalized accelerometer data. \textbf{(b)} Cubic spline interpolation results, enhancing data resolution. \textbf{(c)} Accuracy improvement from combining original and inferred accelerometer samples, with R-squared values indicating goodness of fit. The flowchart depicts the preprocessing, LightGBM-based feature extraction, interpolation, and final SLU model analysis.}
    \label{fig:STAGARCH}
\end{figure*}

To collect sensor readings, we designed a simple Android application that retrieves \texttt{.wav} files from a server hosted on a Raspberry Pi with a static IP. The app plays these audio instructions through a smartphone's loudspeaker while simultaneously recording data from the accelerometer and gyroscope at a 400 Hz sampling rate. The recorded data is then sent back to the server for storage and analysis. This automated setup not only enables remote real-time data analysis—essential for assessing our system’s efficacy and reducing data transfer delays—but also supports parallel data collection using a client-server model. Each device accesses only the necessary metadata, with audio files centrally stored on the Raspberry Pi. The entire experiment was conducted in a soundproof room and lasted 32 continuous hours.

\subsection {Model Selection}
\label{sec:model_selection}

The central role of the machine learning model in the STAG system is to upscale the accelerometer data from its native sampling rate of 200 Hz to the desired 400 Hz range. This enhancement in the sampling rate is vital for increasing both the detail and accuracy of the sensor data. Such an improvement significantly augments the system's acoustic surveillance capabilities, enabling a more precise capture and analysis of sound data, all while operating within the existing permission constraints.

Streaming raw accelerometer and gyroscope data continuously from mobile devices consumes substantial network resources and increases detection risks. Our initial tests revealed that transmitting unprocessed data for 16 hours resulted in 537.6 MB of data, compared to just 1.2 MB for processed data. To improve efficiency and reduce detection risks, we recommend on-device processing, which considerably lowers bandwidth usage by minimizing the data required for transmission, potentially obviating the need for it entirely.

% \begin{figure}[t]
% \centering
% \begin{minipage}{.505\textwidth}
%   \centering
% \captionof{table}{Storage space requirements.}
%   \adjustbox{width=\textwidth}{
%     \begin{tabular}{lll}
%     \toprule
%     \textbf{Raw Data (MB)} & \textbf{Processed data (MB)} & \textbf{Reduction} \\
%     \midrule
%     537.6 & 1.2 & 448 : 1\\
%     \bottomrule
%     \end{tabular}}
% \label{tab:data comp}  
% \end{minipage}%
% \hspace{0.4cm}
% \begin{minipage}{.45\textwidth}
% \captionof{table}{Comparative analysis of machine learning models' memory overhead vs. efficiency.}
% \adjustbox{width=\textwidth}{
% \begin{tabular}{ccc}
% \toprule
% \textbf{Model} & \textbf{Memory Usage (MB)}  & \textbf{Inference time}\\ \midrule
% RNN      &          1149.4           &            0.6335               \\
% CNN      &          1081.1           &            0.6340            \\ 
% Random Forest   &          957              &            0.6383             \\
% % SVM                           &                                      &                         &                         &                                   \\
% \rowcolor{green} LightGBM &  827.7 &    0.6393  \\
% \bottomrule
% \label{tab:model_comp}
% \end{tabular}}
% \end{minipage}
% \end{figure}

\begin{table}[H]
\centering
\caption{Comparative analysis of machine learning models' memory overhead vs. efficiency.}
\begin{tabular}{ccc}
\toprule
\textbf{Model} & \textbf{Memory Usage (MB)}  & \textbf{Inference time (ms)}\\ \midrule
RNN      &          1149.4           &            0.6335               \\
CNN      &          1081.1           &            0.6340            \\ 
Random Forest   &          957              &            0.6383             \\
% SVM                           &                                      &                         &                         &                                   \\
\rowcolor{green} LightGBM &  827.7 &    0.6393  \\
\bottomrule
\label{tab:model_comp}
\end{tabular}
\end{table}

Choosing a machine learning model that requires minimal memory and processing power is essential due to the constrained computational resources and battery life of mobile devices. To identify the optimal model, we evaluated popular algorithms including LSTM, CNN, RNN, Random Forests, and LightGBM, focusing on model complexity, memory usage, processing speed, and accuracy (Table \ref{tab:model_comp}). LightGBM was found to be the most appropriate, offering low memory usage and fast inference times, making it ideal for on-device deployment. Its efficient architecture ensures rapid processing and minimal resource consumption, while its robustness in modeling complex data interactions makes it well-suited for sensor fusion tasks.

\subsection{STAG Architecture}

The STAG system integrates various findings from our research into a unified architecture that enhances sensor data upsampling through a blend of temporal misalignment, sensor data correlation, and machine learning optimization. Outlined in Figure \ref{fig:STAGARCH}, the architecture involves critical stages of Prediction, Interpolation, and Combination, pivotal for precise sensor fusion.

% STAG starts by misaligning accelerometer and gyroscope data sequences, optimizing them through a dual-method upsampling of LightGBM and cubic spline interpolation, with respective R-squared values shown in Figure~\ref{fig:STAGARCH}(a) and (b).

\revised{STAG begins by intentionally misaligning the accelerometer and gyroscope data sequences. These sequences are then optimized using a dual-method upsampling approach employing both LightGBM, which predicts missing data points based on the correlation learned from the training data, and cubic spline interpolation, which smooths the data. The effectiveness of each method is assessed using their respective R-squared ($R^2$) values, which indicate the proportion of variance in the data explained by the model. Higher $R^2$ values signify a better fit to the data. These $R^2$ values are depicted in Figure~\ref{fig:STAGARCH}(a) for LightGBM and Figure~\ref{fig:STAGARCH}(b) for cubic spline interpolation.}
Data accuracy is further refined by merging the prediction and interpolation outputs, achieving an enhanced $R^2$ value depicted in Figure~\ref{fig:STAGARCH}(c).

The processed data is utilized by an SLU model to produce predictive text, transforming misaligned sensor data into actionable insights. This exemplifies STAG’s ability to refine raw data into a format ideal for practical applications.

\section{Evaluation}
\label{sec:eval}
We present a thorough evaluation of STAG, comparing it against established methods like InertiEar \cite{9796890} and StealthyIMU \revised{and accounting for motion artifacts}. We focus on assessing STAG's performance through a structured setup involving data bifurcation, training, testing, and key metrics like WER, SER, and SEER. This detailed analysis aims to validate STAG's effectiveness and highlight its advancements in sensor data analysis for mobile devices.

\subsection{Evaluation Setup}
\begin{figure*}
    \centering
  \includegraphics[width=\textwidth]{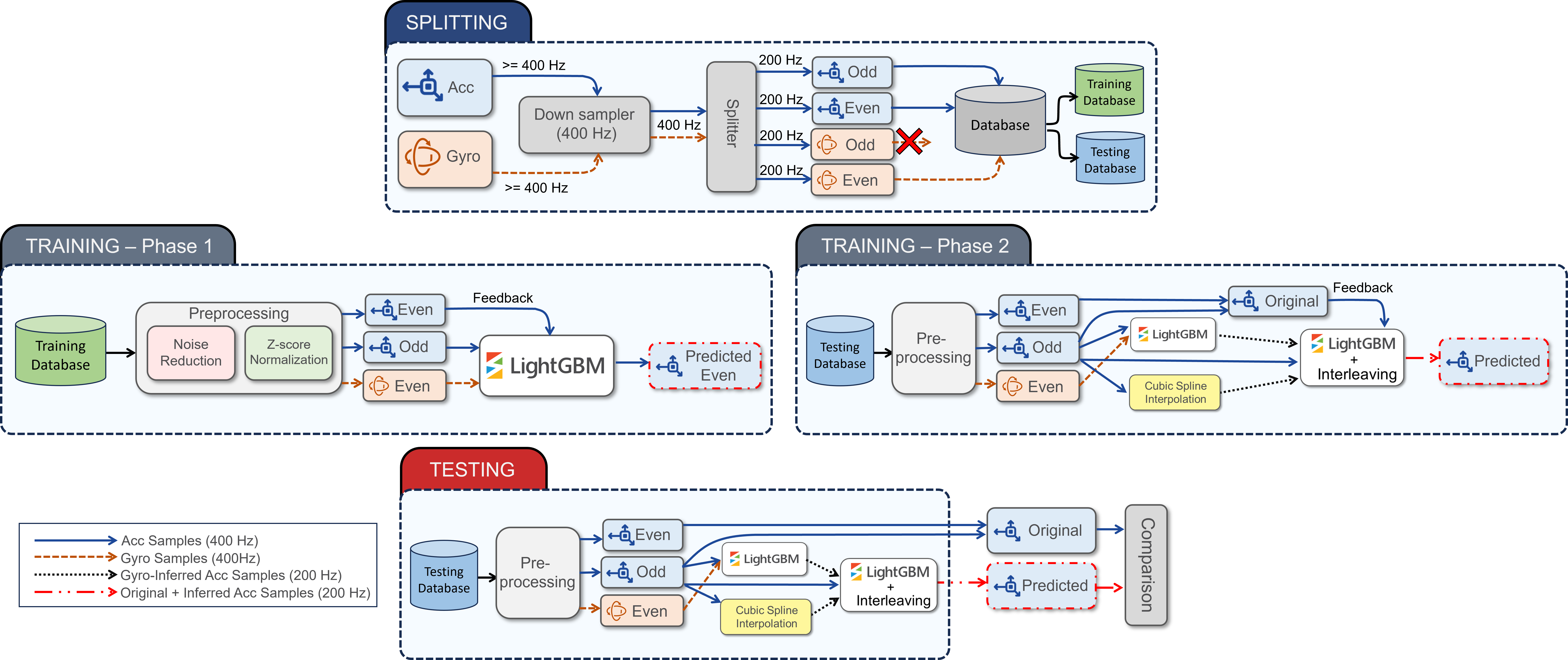}
    \caption{The STAG evaluation process: highlighting data downsampling, bifurcation, and model training/testing phases.}
    \label{fig:setup}
\end{figure*}

Our evaluation of STAG is conducted through a well-defined process, as depicted in Figure~\ref{fig:setup} that provides a comprehensive visual representation of the methodological framework adopted for assessing the system’s effectiveness.

\subsubsection{Down sampling and bifurcation:}
To assess STAG’s predictive accuracy, we first established a ground truth by standardizing the dataset we collected earlier, originally sampled between 400 Hz and 500 Hz, downsampled to 400 Hz. This ensures compatibility with STAG's upscaling capabilities. We then split the sensor readings into alternating (odd and even) samples to emulate an accelerometer operating at a reduced 200 Hz rate, aligning with STAG’s operational scenario. Odd accelerometer and even gyroscope samples were used for STAG input, with even accelerometer samples set aside as feedback for the training model discussed in Section~\ref{sec:training}.

This bifurcation mimics STAG’s enhancement process for an under-sampled accelerometer. The dataset was divided into training (70\%), validation (15\%), and testing (15\%) subsets to optimize and test the LightGBM model’s performance within STAG, ensuring the model learns from complex patterns for precise predictions. Specifically, the dataset consisted of 36,000 samples, with 25,200 samples in the training set, 5,400 samples in the validation set, and 5,400 samples in the test set.

\subsubsection{Training:}
\label{sec:training}
During this phase, the training subset undergoes preprocessing steps like noise reduction and Z-score normalization. These steps prepare the data for processing using the LightGBM algorithm. The training phase is divided into two parts. First, we train a LightGBM model with odd accelerometer values as features and even accelerometer values as feedback. This itself gives us a good baseline, as seen in Figure \ref{fig:STAGARCH}(a). After the model is trained, we carry out cubic spline interpolation on odd accelerometer values, combining the output of the interpolation with the one from the model to train a new combined model. We use mean squared error and R-squared to finetune our model. 

Hyperparameter tuning was performed using a grid search method, combined with 5-fold cross-validation on the training set, to ensure the selected hyperparameters generalized well. The best hyperparameters were chosen based on performance on the validation set.

To mitigate concerns about overfitting with the LightGBM model, we employed careful data partitioning and cross-validation throughout our evaluation, demonstrating the robustness and generalizability of our approach.

\subsubsection{Testing:} The testing subset, subjected to similar preprocessing, is input into the already trained LightGBM model. Here, the algorithm estimates the missing accelerometer samples. These predictions are then merged with the outputs from cubic spline interpolation and further refined through a subsequent application of LightGBM. This dual-phase processing with LightGBM ensures refined accuracy in the system's final output. The final resulting output can be seen in Figure \ref{fig:STAGARCH}(c). The testing resulted in an average R-squared of 0.58.

\subsection{Evaluation Metrics}
To thoroughly assess the performance of STAG, we employ the following metrics:
\begin{itemize}
    \item \textbf{Word Error Rate (WER):} This metric is widely used in automatic speech recognition (ASR) to gauge the accuracy of speech recognition systems. WER calculation is based on the Levenshtein distance, a measure that quantifies the difference between two sequences of letters, which, in this case, represent words in a transcription. Adjustments involving substitution, deletion, or insertion of words are made to align the recognized word sequence with the ground truth. WER is expressed as the ratio of the total number of errors (substitutions, deletions, and insertions) to the total number of words in the standard word sequence \cite{10094784}:
\begin{equation}
WER = \frac{D + S + I}{N} \times 100
\end{equation}
Here, $D$, $S$, $I$, and $N$ denote the number of deletions, substitutions, insertions, and total words in the standard word sequence, respectively.
    \item \textbf{Sentence Error Rate (SER)}: This metric is used to evaluate the accuracy of SLU responses. SER considers a response to be error-free only if all entities in a single voice assistant response are correctly identified. It is a comprehensive measure for evaluating the accuracy of SLU responses and is calculated as  SER is:
\begin{equation}
SER = \frac{\text{\# Incorrect Sentences}}{\text{Total Sentences}}
\end{equation}
    \item \textbf{Single Entity Error Rate (SEER)}: In SLU, a single voice assistant response may contain multiple entities. SEER is akin to WER but focuses on the entities within SLU responses. It measures the accuracy of identifying individual entities as follows:
\begin{equation}
SEER = \frac{\text{\# Incorrect Entities}}{\text{Total Entities}}
\end{equation}
\end{itemize}

These metrics provide a comprehensive framework for evaluating the effectiveness of STAG in both speech recognition and understanding contexts, enabling a holistic assessment of the system's performance.

\subsection{Results}
\label{sec:results}

\begin{table}
    \centering
    
    \caption{Results: STAG achieves 58\% reduction in SER and an 86\% decrease in WER, outperforming the other methods sampling data at 200 Hz.}
   
    \begin{tabular}{llll}
    \toprule
      & \textbf{SER\%} & \textbf{SEER\%} & \textbf{WER\%} \\
    \midrule
    StealthyIMU \cite{inproceedings} & 99.68 & 83.7 & 78.75 \\
    InertiEar \cite{9796890}  & 84.94 & 42 & 24.44 \\
    \rowcolor{green} STAG &  42.83 &  21 &  13.02 \\
    \bottomrule
    \end{tabular}
    \label{tab:results}
\end{table}

\subsubsection{\revised{Comparison with State-of-the-art}}
\revised{STAG's capabilities are compared against two established SLU approaches: \textit{InertiEar} \cite{9796890} and \textit{StealthyIMU} \cite{inproceedings}. Table \ref{tab:results} highlights the key performance metrics, demonstrating STAG's superior accuracy with a WER of 13\% and a SEER of 20\%. These results underscore STAG's effectiveness in transcribing functional words and capturing semantically significant entities like names and locations. Notably, STAG achieved a 58\% reduction in SER and an 86\% reduction in WER compared to data sampled at 200 Hz without up-scaling, as shown in Table \ref{tab:results}.}

\revised{The enhanced performance of STAG can be attributed to its novel approach of inducing temporal misalignment. StealthyIMU, which relies solely on accelerometer data, struggles to recognize complex speech patterns. InertiEar, while using both accelerometer and gyroscope data, uses temporally aligned data. The minimal temporal disparity in InertiEar's aligned data provides no substantial improvement over single-sensor data, leading to suboptimal performance.}

\revised{STAG's strategic temporal misalignment between accelerometer and gyroscope data allows it to exploit the unique characteristics of each sensor type more effectively. As detailed in Section~\ref{subsec: Corr}, STAG leverages the high SNR of the accelerometer's Z-axis and combines it with the temporally misaligned X and Y axes of the gyroscope. This approach optimizes SNR and ensures a more meaningful correlation between data streams, enhancing the accuracy and reliability of the sensor data. Shifting the data streams in time allows STAG to achieve a higher fidelity in reconstructing speech, as evidenced by the superior WER and SEER results. This comparative analysis conclusively underlines the superiority of STAG's methodology. By effectively harnessing the power of temporal misalignment, STAG advances the capabilities of SLU systems, outperforming traditional methods like those employed by InertiEar and StealthyIMU.}

\begin{table}
    \centering
    \caption{\revised{Performance of STAG and StealthyIMU under various motion conditions.}}
    \label{tab:motion_artifacts}
    \begin{tabular}{lc}
    \toprule
    \textbf{Scenario} & \textbf{WER (\%)} \\
    \midrule
    StealthyIMU (Tabletop, Ideal) \cite{inproceedings} & 78.75 \\
    STAG (Tabletop, Ideal) & 13.02 \\
    STAG (Handheld) & 38.02 \\
    STAG (Car, Stationary) & 54.77 \\
    STAG (Car, Moving) & 67.58 \\
    \bottomrule
    \end{tabular}
\end{table}

\subsubsection{\revised{Robustness to Motion Artifacts}}
\revised{To evaluate the robustness of STAG and its resilience to motion artifacts, we tested it under various conditions, as summarized in Table \ref{tab:motion_artifacts}. The optimal scenario for STAG involves the smartphone being placed on a hard, stationary tabletop surface, where external interference vibrations are minimized. For comparison, we also evaluated the state-of-the-art (SOTA) StealthyIMU under the same conditions.}

\revised{In the most common usage scenario, where the smartphone is handheld, we observed a notable increase in the WER by 25\% compared to the ideal case. Additionally, we tested STAG under more challenging conditions, such as when the smartphone was mounted on a car stand, both stationary and while driving over an uneven road with speed bumps. Even under these worst-case scenarios, STAG demonstrated superior performance compared to StealthyIMU.}

\begin{table*}[h!]
\centering
\caption{\revised{End-to-end results: VUI responses, entities detected by STAG, and resulting privacy leaks.}}
\scriptsize % Adjust font size
\renewcommand{\arraystretch}{1.2} % Adjust row spacing
\setlength{\tabcolsep}{4pt} % Adjust column spacing
\begin{tabularx}{\textwidth}{|l|l|X|X|X|c|}
\hline
\textbf{Type} & \textbf{Privacy Entity} & \textbf{User Request} & \textbf{VUI Response} & \textbf{Entities Detected} & \textbf{Privacy Leak} \\ \hline
WEATHER & LOCATION & What's the weather like today? & Today in San Antonio you can expect a high of 87 degrees. There is currently a severe thunderstorm watch in effect. & \{type: 'city', filler: 'memphis'\} & \textcolor{red}{NO} \\ \hline
WEATHER & LOCATION & Is it going to rain at Huntington Beach? & Probably not today in \textbf{Huntington Beach}. There's only a very slight chance of rain. & \{type: 'city', filler: '\textbf{huntington beach}'\} & \textcolor{darkgreen}{YES} \\ \hline
WEATHER & LOCATION & Is there a chance of snow in Chandler? & Tonight in \textbf{Chandler} there's no chance of snow. & \{type: 'city', filler: '\textbf{chandler}'\} & \textcolor{darkgreen}{YES} \\ \hline
REMINDER & TODO/TIME & Add a reminder to pick up prescription & OK, pick up prescription. When do you want to be reminded? & \{type: 'todo', filler: 'pick up; party; supplies'\} & \textcolor{red}{NO} \\ \hline
REMINDER & TODO/TIME & Add a reminder to buy toothpaste & OK, \textbf{buy toothpaste}. When do you want to be reminded? & \{type: 'todo', filler: '\textbf{buy; toothpaste}'\} & \textcolor{darkgreen}{YES} \\ \hline
REMINDER & TODO/TIME & Remind me to do dishes tomorrow at 5 PM & All right, I'll remind you \textbf{tomorrow at 5 PM}. & \{type: 'time', filler: '\textbf{tomorrow; at; five; pm}'\} & \textcolor{darkgreen}{YES} \\ \hline
STOCK UPDATES & SEARCH & What is the stock price of General Electric? & General Electric Company closed down 3.4\% on Friday at \$81.0. & \{type: 'company', filler: 'general motors; company'\} & \textcolor{red}{NO} \\ \hline
STOCK UPDATES & SEARCH & What is the stock price of Berkshire Hathaway? & \textbf{Berkshire Hathaway} Class A closed up 0.75\% on Friday at \$432.00. & \{type: 'company', filler: '\textbf{berkshire hathaway}'\} & \textcolor{darkgreen}{YES} \\ \hline
SUNSET/SUNRISE & LOCATION & When did sunset in Baltimore? & In \textbf{Baltimore} today, the sunset at 4:55 PM. & \{type: 'city', filler: '\textbf{baltimore}'\} & \textcolor{darkgreen}{YES} \\ \hline
SUNSET/SUNRISE & LOCATION & What is the sunrise in Buffalo? & In Buffalo today, sunrise is at 7:01 AM. & \{type: 'city', filler: 'tampa'\} & \textcolor{red}{NO} \\ \hline
AQI INDEX & LOCATION & AQI in Glendale & According to Purple Air, the air quality near the center of Glendale is good with an index ranging from 18 to 26. & \{type: 'city', filler: 'chicago'\} & \textcolor{red}{NO} \\ \hline
AQI INDEX & LOCATION & AQI in Amarillo & According to Air Now, the air quality near the center of \textbf{Amarillo} is good with an index of 21. & \{type: 'city', filler: '\textbf{amarillo}'\} & \textcolor{darkgreen}{YES} \\ \hline
\end{tabularx}

\label{tab:vui_privacy}
\end{table*}

\subsubsection{\revised{End-to-End Attack Scenario}}

\revised{We conducted a series of experiments simulating real-world usage to demonstrate STAG's capability in a realistic end-to-end attack scenario. In our setup, a user interacts with a voice assistant via common VUI requests while STAG, running in the background, captures and processes IMU sensor data. This mirrors a real-world attack where a malicious app could exploit zero-permission sensor access to eavesdrop on user interactions. The experiments focused on common VUI interactions related to weather, reminders, stock updates, sunset/sunrise times, and air quality index (AQI).}

\revised{For each VUI request type, we present the specific user request, the corresponding VUI response, the entities detected by STAG, and whether a privacy leak occurred. As shown in Table \ref{tab:vui_privacy}, STAG successfully identified most of the target privacy entities. These results highlight the potential for STAG to compromise user privacy by exploiting seemingly innocuous VUI interactions.}

% This comparative analysis conclusively underlines the superiority of STAG's methodology. By effectively harnessing the power of temporal misalignment, STAG advances the capabilities of SLU systems, outperforming traditional methods like those employed by \textit{InertiEar} and \textit{StealthyIMU}. The results validate the effectiveness of STAG and highlight its potential to redefine the landscape of sensor-based SLU technologies.

%\revised{The results presented in these evaluations highlight the significant advancements offered by STAG in the field of sensor-based SLU technologies. The superior performance against SOTA methods, coupled with the demonstrated robustness to motion artifacts, validates the effectiveness of STAG's novel approach. By strategically inducing temporal misalignment and effectively fusing accelerometer and gyroscope data, STAG not only overcomes the limitations imposed by current Android security measures but also sets a new benchmark for accuracy and reliability in acoustic eavesdropping.}

% \begin{figure}[htp]
%     \centering
%     \includegraphics[width=8cm]{images/results.png}
%     \caption{results of STAG}
%     \label{fig:arch}
% \end{figure}

% STAG showed remarkable improvements: a 58\% reduction in sentence error rate and an 86\% reduction in word error rate, compared to data sampled at 200 Hz without up-scaling.

\begin{table}[t]
    \centering
    \scriptsize
    \caption{\revised{Prevention strategies for STAG-based attacks.}}
    \label{tab:countermeasures}
    \begin{tabular}{|p{2.1cm}|p{1.9cm}|p{3.5cm}|}
    \toprule
    \textbf{Prevention Strategy} & \textbf{Responsible Party} & \textbf{Practicality} \\
    \midrule
    \makecell{Stricter Access \\ Controls} & Operating System Developers & \cellcolor{green!20} High – Integrates with existing permission systems with minimal additional overhead. \\
    \midrule
    \makecell{Noise Injection \\ Techniques} & Application and OS Developers & \cellcolor{yellow!20} Medium – Requires careful implementation to maintain application performance while obfuscating data. \\
    \midrule
    \makecell{Direct Host \\ Connection for \\ Sensors} & Hardware Manufacturers & \cellcolor{red!20} Low to Medium – May necessitate hardware redesign, increasing development time and costs. \\
    \bottomrule
    \end{tabular}
\end{table}

% \begin{table}
%     \centering
%     \caption{\revised{Motion Artifacts: STAG compared in different day-to-day scenarios}}
   
%     \begin{tabular}{llll}
%     \toprule
%       & \textbf{WER\%} \\
%     \midrule
%     StealthyIMU Tabletop (Ideal)\cite{inproceedings} & 78.75 \\
%     STAG Tabletop (Ideal)  & 13.02 \\
%     STAG Handheld  &  38.02 \\
%     STAG Car Stationary & 54.77 \\
%     STAG Car Moving & 67.58 \\
%     \bottomrule
%     \end{tabular}
%     \label{tab:artifacts}
% \end{table}

\section{Discussion}

\fakepar{Countermeasures} \revised {Several effective countermeasures can be implemented to mitigate STAG-based attacks that exploit zero-permission access to IMUs for eavesdropping. These are summarized in Table~\ref{tab:countermeasures}}. 

\begin{itemize}[leftmargin=1em]
\item {\textbf{Stricter Access Controls:}} \revised{Enforcing stricter access controls on motion sensor data is crucial. Requiring explicit user permissions for accessing accelerometer and gyroscope data, even at lower sampling rates, ensures that only authorized applications can access these sensitive sensors. This significantly reduces the risk of covert speech recognition.}

\item {\textbf{Noise Injection:}} \revised{Introducing noise injection techniques can obfuscate sensor data, hindering attackers' ability to reconstruct accurate speech patterns. By adding benign noise to sensor readings, legitimate applications can still use the data while degrading its usefulness for malicious purposes. This approach balances user privacy with the functionality of authorized applications.}

\item {\textbf{Hardware Modifications:}} \revised{Modifying hardware configurations by connecting the magnetometer and IMU directly to the host system rather than as peripheral devices enhance security. This configuration prevents magnetometer interrupts from affecting the IMU's data buffer (a temporary storage area), thereby isolating sensor data streams and eliminating potential channels for unauthorized data access.}

\end{itemize}

\revised{Implementing these countermeasures significantly diminishes the risk of eavesdropping via STAG-based attacks. This comprehensive strategy provides robust user privacy protection while maintaining the functionality of legitimate, sensor-dependent applications.}

% To prevent the STAG-based zero-permission attack for unauthorized speech recognition, several countermeasures can be implemented. First, imposing stricter access controls on motion sensor data can significantly reduce the risk of exploitation. This includes requiring explicit user permissions for accessing accelerometer and gyroscope data, even at lower sampling rates. This measure ensures that only authorized applications can access sensitive sensor data, thereby protecting user privacy.

% Another effective measure would be to employ noise injection techniques, which add benign noise to sensor readings. This approach obfuscates the data, rendering it less useful for precise speech recognition without significantly degrading the performance of legitimate applications that rely on sensor data.

% Finally, configuring the magnetometer and IMU to be directly connected to the host, rather than as a slave, would help mitigate the issue. In this setup, the magnetometer interrupts would not affect the FIFO buffer within the IMU, as they would be handled by the host instead.

% By addressing these potential vulnerabilities and implementing robust countermeasures, the risk of unauthorized speech recognition via STAG can be significantly mitigated. These measures ensure better protection of user privacy while maintaining the functionality and performance of authorized applications.

\fakepar{Outlook} 
As Android continues to evolve, upcoming versions are likely to include improvements in sensor data handling and synchronization, which could prevent the temporal misalignment exploit utilized by STAG. While our findings reflect the system's present state, they underscore the importance of implementing a timely security patch. By raising awareness of this issue now, we aim to encourage proactive steps to safeguard user privacy.

% The STAG system, while demonstrating robust capabilities, opens exciting avenues for future research to enhance its effectiveness and broad applicability. Currently, validation has been limited to a few mobile device models. Expanding testing across a wider array of hardware configurations will ensure the generalizability of our results and adapt STAG's functionality to diverse technological ecosystems. This broader validation will affirm STAG's versatility and uncover potential optimization areas for various device specifications.

% Additionally, the reliance on the presence of a magnetometer in the device restricts STAG's deployment to specific mobile models. Future research will focus on adapting STAG to work with alternative sensors and configurations, thereby broadening its compatibility and deployment potential. Addressing variations in accelerometer and gyroscope specifications across devices will also enhance STAG's performance and reliability, ensuring consistent results regardless of hardware differences.

The current implementation of STAG leverages the StealthyIMU SLU model, which is optimized for English. This presents an opportunity to integrate more advanced and multilingual SLU models, such as those developed in recent advancements~\cite{wang2023whislu,arora2023universlu}. Incorporating these models will improve accuracy and also extend STAG's utility to multilingual contexts, making it accessible and effective across different cultural and linguistic settings.

Another exciting challenge is optimizing STAG's signal processing algorithms to run on-device, reducing dependency on internet connectivity. By processing data locally, STAG can operate efficiently even in environments with limited connectivity. This optimization will also address concerns related to data storage and transmission, as on-device processing can store only speech recognition outputs, which are significantly smaller than raw sensor data. This approach will enhance user privacy and reduce data storage requirements.

\section{Conclusion}
\label{sec:conclusion}
% What is the big take away from your research. Include any limitations or future work here. 
We introduced STAG, a novel approach designed to circumvent the limitations imposed by Android's 200 Hz sampling rate cap on IMU sensors. By inducing a temporal misalignment between the gyroscope and accelerometer, and leveraging sophisticated data fusion techniques, STAG achieves a significant reduction in word error rate (86\%), demonstrating a robust capability to enhance acoustic surveillance under restricted conditions. The proposed method showcases a technical advancement in exploiting IMU data and highlights the need for more stringent security measures in smartphone sensors.

Our evaluation provides compelling evidence of STAG's ability to manipulate sensor data to overcome existing security protocols, which were thought to mitigate such risks. The approach underscores the persistent vulnerabilities in smartphone sensor security and serves as a critical reminder of the ongoing need for advancements in this area to protect user privacy. %The findings suggest that current measures are insufficient against sophisticated attacks like STAG, calling for a reevaluation of sensor security strategies in modern devices. 

% TO add:
% One whole new paragraph on how the sensor works.
% One new paragraph on sensors in the mobiles and other various sensors used.

\bibliographystyle{ACM-Reference-Format}
\balance
\bibliography{bib}

% % --- Appendix ---%

\end{document}